\documentclass{appolb}
\usepackage{graphicx}

\newcommand{\trento}{T$\mathrel{\protect\raisebox{-2.1pt}{R}}$ENTo}

\begin{document}
\title{There and Sharp Again:  \\ The Circle Journey of Nucleons and Energy Deposition %
\thanks{Presented at the XXIXth International Conference on Ultra-relativistic Nucleus-Nucleus Collisions
(Quark Matter 2022)}%
}
\author{Giuliano Giacalone
\address{Institut f\"ur Theoretische Physik, Philosophenweg 16, 69120 Heidelberg, Germany}
\\[3mm]
}
\maketitle
\begin{abstract}
A central question in high-energy nuclear phenomenology is how the geometry of the quark-gluon plasma (QGP) formed in relativistic nuclear collisions is precisely shaped. In our understanding of such processes, two features are especially crucial for the determination of the QGP geometry, respectively, the nucleon size and the energy deposition scheme. This contribution reports on the (circular) evolution of such features in state-of-the-art model incarnations of heavy-ion collisions over the past seven years. Ideas for future directions of investigation are pointed out.
\end{abstract}

\noindent \textbf{1. Nucleon (or sub-nucleon) size and energy deposition.}~ In the state-of-the-art picture of a heavy-ion collision, the interaction process acts a quantum measurement for the transverse positions of the constituents of the colliding nuclei. A nucleus viewed in the laboratory frame is associated, then, with a two-dimensional profile representing a snapshot of its Lorentz-boosted content at the time of interaction. The profile for, say, nucleus $A$, is of the form (${\bf x}$ is a transverse coordinate):
\begin{equation}
\label{eq:tA_nucl}
    T_A ({\bf x}) = \sum_{j=1}^{A_A}  \lambda_j g({\bf x}; {\bf x}_j w), \hspace{10pt} g({\bf x}; {\bf x}_j w) = \frac{1}{2\pi w^2} \exp \left ( - \frac{({\bf x} - {\bf x}_j)^2}{2 w^2} \right ),
\end{equation}
where $j$ labels the $A_A$ nucleons in nucleus $A$, $\lambda_j$ is a random fluctuation associated with nucleon $j$, and $g({\bf x}; {\bf x}_j, w)$ is the nucleon profile, commonly taken as a Gaussian where ${\bf x}_j$ represents the center position of nucleon $j$ within nucleus $A$, and $w$ is the anticipated nucleon size.

What is the appropriate nucleon size for high-energy scattering? We know little about it. In deep inelastic scattering (DIS) at low $x$, the dipole cross section at a given impact parameter is of the form $\frac{d\sigma_{q\bar q}}{d^2{\bf b}}\propto r^2 \alpha_s xg(\mu^2,x) T({\bf b})$, where $T({\bf b})$ weights the gluon density depending on the distance from the target's center. The data-driven analysis by Caldwell and Kovalsky \cite{Caldwell:2010zza} based on the diffractive production of $J/\Psi$ at HERA concludes that $T({\bf b})$ can be modeled as a 2D Gaussian with width $0.35$ fm, corresponding to an rms radius in two dimensions, dubbed the \textit{two-gluon radius}, of $\sqrt{2}w=0.50$ fm. In 2012, this $w$ has been used as input in the IP-Glasma model by Schenke, Tribedy, and Venugopalan \cite{Schenke:2012wb}, based on the color glass condensate (CGC) effective theory of high-energy QCD \cite{Gelis:2010nm}. 

In a description of high-energy collisions, it is appropriate to consider as well the inner structure of the colliding nucleons. One introduces an effective parton structure by replacing Eq.~(\ref{eq:tA_nucl}) with
\begin{equation}
\label{eq:tA_sub}
    T_A({\bf x}) = \sum_{j=1}^{A_{\rm A}} \sum_{q=1}^{n_c} \lambda_q g(x; {\bf x}_q, w_q),
\end{equation}
where now for each nucleon, $j$, one samples the centers, ${\bf x}_q$, of $n_c$ constituents from a Gaussian distribution of width $w$, and treats each sampled constituent as a Gaussian profile of width $w_q$. Some insights on the number of constituents and their widths can be obtained as well from diffractive production in DIS, by studying the incoherent J/$\Psi$ production, which probes fluctuations in the proton content. The data-driven analysis of Mäntysaari and Schenke \cite{Mantysaari:2016ykx} in the CGC framework infers three constituents of width $w_q=0.11$ fm. Note that the number of constituents is poorly constrained, and may be higher \cite{Mantysaari:2022ffw}. A nucleon size of about $0.4$ fm and three constituents of size $0.11$ fm are implemented in the latest version of IP-Glasma \cite{Schenke:2020mbo}.

The second crucial ingredient for the geometry of the QGP is the energy deposition scheme. This refers to some function of $T_A({\bf x})$ and $T_B({\bf x})$ that turns them into an energy density in the transverse plane, corresponding to the initial condition for the subsequent QGP formation at proper time $\tau=0^+$. For high-energy nuclear scattering insights about the energy deposition come again from the CGC framework. A robust prediction of such a framework (which is nowadays textbook material \cite{Gelis:2019yfm}) is that the average energy density in the scattering of two CGCs at $\tau=0^+$ is of the form
\begin{equation}
    \langle T^{00} ({\bf x}) \rangle [{\rm GeV}/{\rm fm}^3] \propto Q_A^2({\bf x}) Q_B^2 ({\bf x}).
\end{equation}
The saturation scales, $Q^2_{A,B}$, involve the same function $T({\bf b})$ appearing in the dipole cross section, and as such are essentially proportional to $T_{A,B}$.\footnote{Note that in \trento{} the function $T_A$ includes only the nucleons that participate in the collisions, whereas all nucleons are considered in the nuclear profiles in IP-Glasma.} The IP-Glasma model is based on such a scaling, as confirmed by comparisons with the IP-Jazma parametrization \cite{Nagle:2018ybc,Snyder:2020rdy}, where in every collision event one takes $T^{00} \propto T_AT_B$. 

Features of the CGC aside, one can confidently state that, up to quantitative corrections, experimental evidence in heavy-ion collisions points to an Ansatz of the kind $(T_A T_B)^m$ for the initial energy density of the system. Note that the random normalizations, $\lambda_j$ in Eq.~(\ref{eq:tA_nucl}), and $\lambda_q$ in Eq.~(\ref{eq:tA_sub}), are less important for the QGP geometry compared to $w$, $w_q$, or the energy deposition scheme, and will not be part of the present discussion.

\noindent \textbf{2. 2015 -- Entering the precision era.}~  
At the end of 2014, Bass, Bernhard, and Moreland (hereafter the Duke group), devise a new Ansatz \cite{Moreland:2014oya} for the entropy density at the beginning of the hydrodynamic phase of the QGP, which takes the form of a generalized average (\trento{} model):
\begin{equation}
\label{eq:trento}
    \frac{dS}{dy}(\tau)~ [1/{\rm fm}^2] \equiv \tau s({\bf x},\tau) \propto \left ( \frac{T_A^p + T_B^p}{2} \right)^{1/p}
\end{equation}
If one sets $p=0$, the entropy density becomes $dS/dy \propto \sqrt{T_AT_B}$. Note that this is the only combination of the form $(T_A T_B)^m$ that can result from Eq.~(\ref{eq:trento}). Let us reformulate this $p=0$ model in a slightly different way. We introduce the energy density per unit rapidity at the initial time, and assume it is proportional to $(T_AT_B)^{2/3}$,
\begin{equation}
    \frac{dE}{dy} (\tau = 0^+) {\rm [GeV/fm^2]}  \equiv \lim_{\tau \to 0^+} \tau e({\bf x},\tau) \propto \left (T_A T_B \right )^{2/3}, 
\end{equation}
where $e({\bf x},\tau)$ is the energy density of the system in units GeV/fm$^3$. The initial $dE/dy$ can, then, be evolved (e.g., freely streamed) for a short time scale (e.g., $0.4$ fm/$c$), to obtain the energy density at finite $\tau,$ $e({\bf x},\tau)$. The latter retains the geometric features of the initial $dE/dy$. As such, upon application of the equation of state of high-temperature QCD, $s \propto e^{3/4}$, the resulting entropy per unit rapidity, $dS/dy$, will follow $(T_AT_B)^{2/3 \times 3/4 = 1/2}$, as in the $p=0$ Ansatz.

The Duke group recognizes that this model provides geometric features for the initial profiles that are remarkably close to those calculated within IP-Glasma, in particular regarding the eccentricities, $\varepsilon_n$, which are the seeds of the anisotropic flow coefficients, $v_n$. The breakthrough of the \trento{} prescription is that it allows one to produce results of potentially similar quality as those obtained in IP-Glasma within a much lighter computational framework. The possibility of mass-producing millions of \trento{} initial conditions opens a new way to compare initial-state calculations to experimental data. In particular, one can precisely assess observables that are too costly for hydrodynamic simulations, such as higher-order fluctuations of $v_n$ \cite{Giacalone:2017uqx}, to scrutinize in great detail the role of the hydrodynamic response ($v_n \propto \varepsilon_n$). Thanks to \trento{}, the phenomenology of the soft sector of heavy-ion collisions enters a new era of precision.

\begin{figure}
\centerline{%
\includegraphics[width=.9\linewidth]{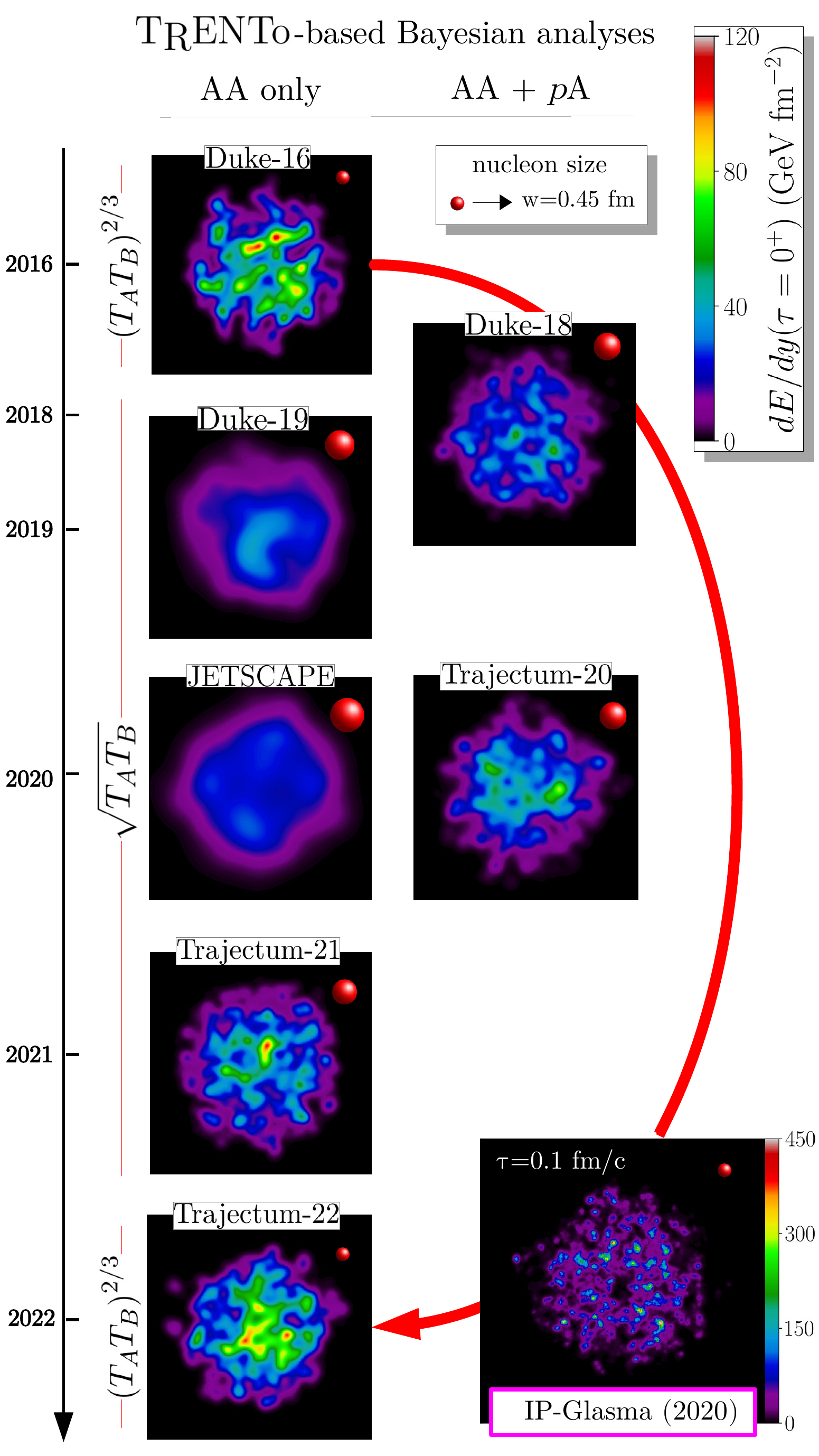}}
\caption{Circle trip of the initial energy density per unit rapidity, $dE/dy$, and of the nucleon size, $w$, in initial-state parametrizations inferred from Bayesian analyses of nucleus-nucleus and proton-nucleus collisions in the past six years. Each plot shows one Pb+Pb collision at $b=0$ and $\sqrt{s_{\rm NN}}=2.76$ TeV. Each profile in the $(x,y)$ plane is plotted in the square $-10 \leq x \leq 10$ fm, $-10 \leq y \leq 10$ fm.}
\label{fig:1}
\end{figure}

\noindent \textbf{3. 2016 -- The first Bayesian analysis.}~ In 2016, the Duke group performs the first Bayesian analysis (hereafter referred to as Duke-16) of 2.76 TeV Pb+Pb data \cite{Bernhard:2016tnd}. The analysis aims at the inference of high-probability (or Maximum A Posteriori, MAP) parameters for the initial state and the transport properties of the QGP. The \trento{} model is here used as an entropy density at the beginning of hydrodynamics ($\tau=0.4$ fm/$c$). The analysis is very successful. Well-defined posterior distributions of parameters are obtained, their correlations are quantified for the first time, and the first data-driven extraction of $\eta/s(T)$ is achieved. The most notable result is however the fact that the posterior distribution of the generalized mean parameter, $p$, is strongly peaked around $p=0$, confirming an initial geometry in agreement with the results of IP-Glasma. Another important result concerns the nucleon size. The prior range for this parameter is chosen between 0.4 fm and 1 fm. The MAP value is around $w=0.45$ fm, meaning that data gives strong preference for small values of the nucleon size.  Remarkably, then, both analyses of diffractive J/$\Psi$ production at HERA and a Bayesian analysis of heavy-ion collisions based on the \trento{} model return nucleons of the same size. Probably, this is not just a happy coincidence. As of 2016, the energy deposition is essentially $dE/dy (\tau=0^+) \propto (T_A T_B)^{2/3}$, with $w=0.45$ fm. See Fig.~\ref{fig:1}.

\noindent \textbf{4. 2019/2020 -- Prescription changes and nucleons swell.}~Things take a dramatic turn with a new Bayesian analysis of the Duke group (Duke-19) \cite{Bernhard:2019bmu}, inferring more parameters from data (e.g., the temperature dependence of the specific bulk viscosity, $\zeta/s$), and including a free-streaming pre-hydrodynamic phase. The \trento{} model becomes the initial condition for this free streaming phase and plays the role of the energy density per unit rapidity, $dE/dy$. Unsurprisingly, the found MAP value of $p$ remains $p=0$, i.e., the only allowed combination of the type $(T_AT_B)^m$. It should be noted that moving from $dS/dy$ (Duke-16) to $dE/dy$ yields a more diffuse profile and smaller local energy density fluctuations. The global analysis is nonetheless able to provide an excellent description of the considered experimental data, although with one issue. The nucleon size, still constrained from a prior range $w\in[0.4,1]$ fm, acquires now a MAP value $w=0.96$ fm, doubling the 2016 estimate.  The combined effect of this large nucleon size and more diffuse energy density deposition leads to extremely smooth initial conditions for the QGP, as shown in Fig.~\ref{fig:1}. This result has been found as well at the end of 2020 by the JETSCAPE collaboration in their Bayesian analysis of Pb+Pb and Au+Au data \cite{JETSCAPE:2020mzn}, with a broader prior range for the nucleon size parameter. Figure~\ref{fig:1} shows, in particular, the JETSCAPE initial condition obtained from MAP parameters employing so-called Grad-type viscous corrections at freeze-out, where the MAP width is $w=1.12$ fm. A consequence of these smooth profiles is the damping of pressure gradients in the fluid, which in turn require much reduced $\zeta/s$ ($(\zeta/s)_{\rm max}\approx0.03$ in Duke-19, while $(\zeta/s)_{\rm max}\approx0.1$ with IP-Glasma initial conditions) to reproduce the measured radial flow. Note that similar results have been obtained as well in the recent analysis of Parkkila \textit{et al.} \cite{Parkkila:2021tqq,Parkkila:2021yha} ($w\approx0.8$ fm).

We remark, then, that the latest version of the IP-Glasma model of initial condition also appears in 2020 \cite{Schenke:2020mbo}. As shown in Fig.~\ref{fig:1}, the predicted initial profile is sharp and lumpy, due to the fine nucleon structure. At the end of 2020, IP-Glasma and \trento{} models based on Bayesian analyses of nucleus-nucleus data are starkly inconsistent, both in terms of initial profiles and in terms of implemented bulk viscosities.

Let us comment, then, on the Bayesian analysis of Ref.~\cite{Moreland:2018gsh} (Duke-18) which uses the same model as in Duke-19, albeit including experimental data from proton-nucleus collisions. Following the inclusion of the nucleon sub-structure in IP-Glasma \cite{Mantysaari:2017cni}, the Duke-18 calculation introduces sub-nucleonic constituents in agreement with Eq.~(\ref{eq:tA_sub}) to make the \trento{} Ansatz viable for proton-nucleus collisions. The notable result is that the $p=0$ Ansatz remains the only viable model as well when $p$+Pb data is included. This is probably driven by the $A$-$A$ data, as it is not known at present whether analyzing $p$-$A$ alone would give the same result. Similar results are obtained in the first Bayesian analysis performed within the Trajectum framework (Trajectum-20) \cite{Nijs:2020ors}. Both Duke-18 and Trajectum-20 infer rather large nucleon sizes, $w\approx0.9$ fm, though with a constituent size $w_q\approx0.5$ fm. Comparing, then, these profiles with the JETSCAPE and Duke-19 ones in Fig.~\ref{fig:1}, we can see the additional short-range structures coming from the nucleon constituents. Note that this additional structure does not imply larger values of $\zeta/s$, which remains very close to zero \cite{Nijs:2020ors}.

\noindent \textbf{5. 2021/2022 -- Deflating the quark-gluon plasma via $\boldmath{\zeta/s}$.}~The profiles of Duke-19 or JETSCAPE are clearly questionable.  There is no apparent issue with an initial energy density of the form $(T_A T_B)^{1/2}$, but too large size of nucleons seems problematic. Although we may not fully understand why experimental data tends to favor such smooth initial profiles, it may be possible to find observables that depend in a dramatic way on the nucleon size, allowing us to stress-test these results. 

The first such observable has been pointed out by Giacalone, Schenke, and Shen \cite{Giacalone:2021clp}. It is the Pearson correlation between the charged-hadron mean transverse momentum, $\langle p_t \rangle$, and anisotropic flow, $v_n^2$, denoted by $\rho(\langle p_t \rangle,v_n^2)$ \cite{Bozek:2016yoj}. As emphasized by the ALICE collaboration \cite{ALICE:2021gxt}, this correlator presents a unique model dependence. For instance, $\rho(v_3^2,\langle p_t \rangle)$ is negative at almost all centralities in the JETSCAPE results, while it is always positive with IP-Glasma initial conditions, in agreement with experimental measurements \cite{ATLAS:2019pvn,ALICE:2021gxt,ATLAS:2022dov}. This largely comes from the huge difference in the implemented size of nucleons. Even the model outcome of the sophisticated Bayesian analysis performed in 2021 within the Trajectum framework (Trajectum-21) by Nijs and van der Schee \cite{Nijs:2021clz}, with the well-structured initial profile shown in Fig.~\ref{fig:1}, yields hydrodynamic results for $\rho(v_n^2,\langle p_t \rangle)$ that are inconsistent with data \cite{ATLAS:2022dov}. The reason is partly that the nucleon size remains too large, of order 0.85 fm.

One way to fix this issue is to look at the total inelastic nucleus-nucleus cross section, $\sigma_{AA}$. In \trento{} and IP-Glasma, the nucleon size determines the probability of interaction between two ions at a given impact parameter. As the total cross section is constrained fairly well by Glauber fits of multiplicity distributions, it provides an experimental handle on $w$. The latest Trajectum study (Trajectum-22) \cite{Nijs:2022rme} addresses this problem by essentially enforcing the Bayesian analysis to return the correct value of $\sigma_{AA}$. Not only the nucleon size shrinks to a value close to 0.5 fm, but the exploration of a doubly-generalized average with an additional parameter, $q$,
\begin{equation}
  \frac{dE}{dy} (\tau=0^+) \propto  \left (  \frac{T_A^p + T_B^p}{2} \right )^{q/p},
\end{equation}
shows that, when the $\sigma_{AA}$ constraint is properly considered and the computed $\rho(v_n^2,\langle p_t \rangle)$ correlators are qualitatively consistent with data, the favored initial-state model has $q\approx4/3$ and $p\approx0$, i.e., $dE/dy \propto (T_BT_B)^{2/3}$ \cite{prep}. The circle is closed (see Fig.~\ref{fig:1}). We are back to the nucleon size and energy deposition of Duke-16 (albeit with four constituents per nucleon, and $w_q\approx0.4$ fm). Moreover, the sharper profile does now impact visibly the extracted $\zeta/s(T)$, whose maximum increases by a large factor \cite{Nijs:2022rme}.

\noindent \textbf{6. What now?}~ Recapping, as of 2022, the state-of-the-art initial state of the quark-gluon plasma returned by Bayesian analyses contains an initial $dE/dy$ that scales like $(T_AT_B)^{2/3}$, with a nucleon size around $w\approx0.5$ fm, and 3-4 constituents per nucleon having $w_q\approx0.4$ fm. It is yet to be understood how this model is precisely connected to IP-Glasma, although their predictions appear to be reasonably consistent. 

To improve the state-of-the-art, one obvious suggestion is that of including in future Bayesian analyses multi-particle correlation observables that offer a stronger sensitivity to the initial condition. The natural candidates are the $\rho(v_n^2,\langle p_t \rangle)$ correlators, though much insight would would come as well from the relative fluctuation of $v_3$, $v_3\{4\}/v_3\{2\}$, or the skewness of the fluctuations of $\langle p_t \rangle$ \cite{Giacalone:2017uqx,Giacalone:2020lbm}. At the price of increasing dramatically the computational effort, these observable would indeed permit global analyses to yield much tighter constraints on initial-state properties.

Second, we have to understand in greater detail the fundamental origin of the strong dependence of observables on model features. One way to do this is to supplement event-by-event calculations with a model-independent description of the initial states in terms of correlation functions of fluctuating fields. Techniques to do so have been discussed in the past \cite{Floerchinger:2013rya,Blaizot:2014nia}. Does $\rho(v_n^2,[p_t])$ depend strongly on the nucleon size because the latter modifies the local average of the energy density field, or because it modifies the local variance? What is the main difference between implementing $w_q=0.4$ fm as in Trajectum-22 and $w_q=0.11$ fm as in IP-Glasma? Is it in the local variance of the density field, or in its correlation length?

One final suggestion about the inclusion in Bayesian analyses of features of the nucleus lead-208 itself. Introducing a nuclear skin thickness parameter would be especially interesting, as this feature affects $\sigma_{AA}$ and the sharpness of the average density profiles. The nucleon density in a nucleus, $\rho_m$, can be written as $\rho_m=\rho_p +\rho_n$, where $\rho_{n(p)}$ is the density of neutrons(protons). While $\rho_p$ is known from low-energy experiments, $\rho_n$, and in particular its skin, is not. One could attempt, thus, at inferring the neutron density of $^{208}$Pb via the Bayesian analysis of high-energy data. This would in turn provide an independent estimate of the neutron skin of this nucleus, a hot topic in low-energy nuclear physics due to its relevance for the understanding of the properties of neutron stars \cite{PREX:2021umo,Reed:2021nqk}.

\noindent \textbf{7. Acknowledgment.}~ The author expresses his gratitude to all the colleagues that across the year 2021 have helped make progress on the issue of the nucleon size in the results of Bayesian analyses. I acknowledge in particular Bjoern Schenke, Chun Shen, Wilke van der Schee, Govert Nijs, Matt Luzum, Jean-Yves Ollitrault, Jamie Nagle, Jiangyong Jia, You Zhou for fruitful discussions on this matter. This research work is funded by the Deutsche Forschungsgemeinschaft (DFG, German Research Foundation) under Germany's Excellence Strategy EXC2181/1-390900948 (the Heidelberg STRUCTURES Excellence Cluster), within the Collaborative Research Center SFB1225 (ISOQUANT, Project-ID 273811115).


\small 
\begin{thebibliography}{99}

\footnotesize

\bibitem{Caldwell:2010zza}
A.~Caldwell and H.~Kowalski,
Phys. Rev. C \textbf{81}, 025203 (2010)
doi:10.1103/PhysRevC.81.025203

\bibitem{Schenke:2012wb}
B.~Schenke, P.~Tribedy and R.~Venugopalan,
Phys. Rev. Lett. \textbf{108}, 252301 (2012)
doi:10.1103/PhysRevLett.108.252301

\bibitem{Gelis:2010nm}
F.~Gelis, E.~Iancu, J.~Jalilian-Marian and R.~Venugopalan,
Ann. Rev. Nucl. Part. Sci. \textbf{60}, 463-489 (2010)
doi:10.1146/annurev.nucl.010909.083629

\bibitem{Mantysaari:2016ykx}
H.~M\"antysaari and B.~Schenke,
Phys. Rev. Lett. \textbf{117}, no.5, 052301 (2016)
doi:10.1103/PhysRevLett.117.052301

\bibitem{Mantysaari:2022ffw}
H.~M\"antysaari, B.~Schenke, C.~Shen and W.~Zhao,
Phys. Lett. B \textbf{833}, 137348 (2022)
doi:10.1016/j.physletb.2022.137348

\bibitem{Schenke:2020mbo}
B.~Schenke, C.~Shen and P.~Tribedy,
Phys. Rev. C \textbf{102}, no.4, 044905 (2020)
doi:10.1103/PhysRevC.102.044905


\bibitem{Gelis:2019yfm}
F.~Gelis,
Cambridge University Press, 2019,
ISBN 978-1-108-48090-1, 978-1-108-57590-4

\bibitem{Nagle:2018ybc}
J.~L.~Nagle and W.~A.~Zajc,
Phys. Rev. C \textbf{99}, no.5, 054908 (2019)
doi:10.1103/PhysRevC.99.054908

\bibitem{Snyder:2020rdy}
R.~Snyder, M.~Byres, S.~H.~Lim and J.~L.~Nagle,
Phys. Rev. C \textbf{103}, no.2, 024906 (2021)
doi:10.1103/PhysRevC.103.024906

\bibitem{Moreland:2014oya}
J.~S.~Moreland, J.~E.~Bernhard and S.~A.~Bass,
Phys. Rev. C \textbf{92}, no.1, 011901 (2015)
doi:10.1103/PhysRevC.92.011901

\bibitem{Giacalone:2017uqx}
G.~Giacalone, J.~Noronha-Hostler and J.~Y.~Ollitrault,
Phys. Rev. C \textbf{95}, no.5, 054910 (2017)
doi:10.1103/PhysRevC.95.054910

\bibitem{Bernhard:2016tnd}
J.~E.~Bernhard, J.~S.~Moreland, S.~A.~Bass, J.~Liu and U.~Heinz,
Phys. Rev. C \textbf{94}, no.2, 024907 (2016)
doi:10.1103/PhysRevC.94.024907

\bibitem{Bernhard:2019bmu}
J.~E.~Bernhard, J.~S.~Moreland and S.~A.~Bass,
Nature Phys. \textbf{15}, no.11, 1113-1117 (2019)
doi:10.1038/s41567-019-0611-8

\bibitem{JETSCAPE:2020mzn}
D.~Everett \textit{et al.} [JETSCAPE],
Phys. Rev. C \textbf{103}, no.5, 054904 (2021)
doi:10.1103/PhysRevC.103.054904

\bibitem{Parkkila:2021tqq}
J.~E.~Parkkila, A.~Onnerstad and D.~J.~Kim,
Phys. Rev. C \textbf{104}, no.5, 054904 (2021)
doi:10.1103/PhysRevC.104.054904

\bibitem{Parkkila:2021yha}
J.~E.~Parkkila, A.~Onnerstad, F.~Taghavi, C.~Mordasini, A.~Bilandzic and D.~J.~Kim,
[arXiv:2111.08145 [hep-ph]].

\bibitem{Moreland:2018gsh}
J.~S.~Moreland, J.~E.~Bernhard and S.~A.~Bass,
Phys. Rev. C \textbf{101}, no.2, 024911 (2020)
doi:10.1103/PhysRevC.101.024911

\bibitem{Mantysaari:2017cni}
H.~M\"antysaari, B.~Schenke, C.~Shen and P.~Tribedy,
Phys. Lett. B \textbf{772}, 681-686 (2017)
doi:10.1016/j.physletb.2017.07.038

\bibitem{Nijs:2020ors}
G.~Nijs, W.~van der Schee, U.~G\"ursoy and R.~Snellings,
Phys. Rev. Lett. \textbf{126}, no.20, 202301 (2021)
doi:10.1103/PhysRevLett.126.202301

\bibitem{Giacalone:2021clp}
G.~Giacalone, B.~Schenke and C.~Shen,
Phys. Rev. Lett. \textbf{128}, no.4, 042301 (2022)
doi:10.1103/PhysRevLett.128.042301

\bibitem{Bozek:2016yoj}
P.~Bozek,
Phys. Rev. C \textbf{93}, no.4, 044908 (2016)
doi:10.1103/PhysRevC.93.044908

\bibitem{ALICE:2021gxt}
S.~Acharya \textit{et al.} [ALICE],
[arXiv:2111.06106 [nucl-ex]].

\bibitem{ATLAS:2019pvn}
G.~Aad \textit{et al.} [ATLAS],
Eur. Phys. J. C \textbf{79}, no.12, 985 (2019)
doi:10.1140/epjc/s10052-019-7489-6

\bibitem{ATLAS:2022dov}
 ATLAS Collaboration,
[arXiv:2205.00039 [nucl-ex]].

\bibitem{Nijs:2021clz}
G.~Nijs and W.~van der Schee,
[arXiv:2110.13153 [nucl-th]].

\bibitem{Nijs:2022rme}
G.~Nijs and W.~van der Schee,
[arXiv:2206.13522 [nucl-th]].

\bibitem{prep}

G.~Nijs and W.~van der Schee, to appear

\bibitem{Floerchinger:2013rya}
S.~Floerchinger and U.~A.~Wiedemann,
Phys. Lett. B \textbf{728}, 407-411 (2014)
doi:10.1016/j.physletb.2013.12.025

\bibitem{Blaizot:2014nia}
J.~P.~Blaizot, W.~Broniowski and J.~Y.~Ollitrault,
Phys. Lett. B \textbf{738}, 166-171 (2014)
doi:10.1016/j.physletb.2014.09.028

\bibitem{Giacalone:2020lbm}
G.~Giacalone, F.~G.~Gardim, J.~Noronha-Hostler and J.~Y.~Ollitrault,
Phys. Rev. C \textbf{103}, no.2, 024910 (2021)
doi:10.1103/PhysRevC.103.024910

\bibitem{PREX:2021umo}
D.~Adhikari \textit{et al.} [PREX],
Phys. Rev. Lett. \textbf{126}, no.17, 172502 (2021)
doi:10.1103/PhysRevLett.126.172502

\bibitem{Reed:2021nqk}
B.~T.~Reed, F.~J.~Fattoyev, C.~J.~Horowitz and J.~Piekarewicz,
Phys. Rev. Lett. \textbf{126}, no.17, 172503 (2021)
doi:10.1103/PhysRevLett.126.172503

\end{thebibliography}
\end{document}